\documentclass[aps,prc,preprint,longbibliography,showkeys]{revtex4-1} 

\usepackage{hyperref}
\usepackage{amsmath}  
\usepackage{amsfonts} 
\usepackage{graphicx} 
\usepackage{subfigure}
\usepackage{color}
\begin{document}

\title{Surface plasmon resonance for in-plane
  birefringence measurement of anisotropic thin organic film}

\author{Amrit Kumar}
\affiliation{Department of Physics, Birla Institute of Technology and
  Science, Pilani (BITS Pilani), 333031, India.}

\author{Raj Kumar Gupta}
\email{raj@pilani.bits-pilani.ac.in}
\affiliation{Department of Physics, Birla Institute of Technology and
  Science, Pilani (BITS Pilani), 333031, India.}

\author{Manjuladevi V}
\affiliation{Department of Physics, Birla Institute of Technology and
  Science, Pilani (BITS Pilani), 333031, India.}

\author{Ashutosh Joshi}
\affiliation{Department of Physics, Birla Institute of Technology and
  Science, Pilani (BITS Pilani), 333031, India.}

\date{\today}

\begin{abstract}
  The measurement of in-plane birefringence ($\Delta{n}$) of ultrathin
  film is challenging due to a significant deviation of physical
  properties of materials in ultrathin regime as compared to that in
  bulk state.  Surface plasmon resonance (SPR) phenomenon can be
  employed to measure change in refractive index of ultrathin film at
  a very high resolution. This article discusses simulation of SPR
  phenomenon in Kretschmann configuration for the measurement of
  $\Delta{n}$ in organic thin film exhibiting
  nematic-like ordering on the two dimensional gold surface. The
  distribution of plasmonic field on the gold surface was found to be
  anisotropic. This suggested that the coupling plasmonic field with
  that of organic thin film exhibiting nematic-like ordering on the
  gold surface will be non-isotropic. Therefore, a non-zero difference
  in resonance angle (RA) was obtained from SPR measurement performed
  along the optic-axis (OA) and orthogonal to OA of the in-plane
  nematic ordering ($\Delta\theta$). A calibration surface showing the variation of
  ($\Delta\theta$) as a function of $\Delta{n}$ and thickness of thin
  organic film consisting of shape anisotropic tilted molecules
  exhibiting nematic-like ordering on gold surface was obtained. This
  calibration surface was employed for the measurement of $\Delta{n}$
  of single layer of Langmuir-Blodgett films of cadmium stearate
  (CdSA) and 4'-octyl-4-biphenylcarbonitrile (8CB) deposited on SPR
  chips. The thickness of the LB films was estimated using X-ray
  reflectivity measurement and $\Delta\theta$ was measured using a
  home built SPR instrument. The $\Delta{n}$ values were found to be
  0.012 and 0.022 for ultrathin films of CdSA and 8CB molecules,
  respectively.
  
\end{abstract}


\maketitle

\section*{Introduction}
The optical phenomenon surface plasmon resonance (SPR) is very popular
owing to its remarkable application in the field of sensors. The
phenomenon facilitates a highly sensitive and label free sensing for a
variety of biological and chemical
analytes~\cite{Homola2008,Wu2019,Gupta2017}. The underlying principle
for a SPR sensor is based on measurement of changes in refractive
index (RI) at a very high resolution due to molecular
interactions. The surface plasmon polaritons (SPP) can be excited at a
metal-dielectric interface by an incident electromagnetic wave
traveling via a coupling medium with RI greater than 1.0.  The
resonance can be established by matching the wavevectors of the
incident and the SPP waves. At the resonance, a maximum energy will
transfer from the incident wave to the SPP wave leading to extinction
of the characteristic incident electromagnetic wave from the
spectrum~\cite{Homolabook2006,Abdulhalim2008,Prabowo2018}. In the
widely utilized Kretschmann configuration of SPR, a p-polarized
monochromatic electromagnetic wave is allowed to incident on the metal
surface via a coupling prism~\cite{Krets1968,Krets1971}. In order to
establish the SPR, the angle of incidence is varied and the reflected
intensity is recorded. At resonance, the reflected intensity
diminishes to minimum. The resonance angle is unique for the given
metal-dielectric interface.  Therefore, any adsorption of analytes at
the metal-dielectric interface during sensing can alter the dielectric
nature and hence resonance angle (RA) shifts. The shift in RA can be
measured very precisely and the corresponding change in RI can be
calculated theoretically using the Fresnel's relations~\cite{devSNB}.
In addition to the traditional sensing applications, the SPR
phenomenon can also be used for the measurement of optical anisotropy
in thin films~\cite{devJMS}, temperature
measurement~\cite{SPR_temp,SPRtemp2,Lu2016} and optical
filter~\cite{wangAPL}. A typical resolution of the Kretschmann
configuration based SPR instrument lies in the range of $10^{-5}$ to
$10^{-7}$~RIU~\cite{devSNB,resolution1,Homola_rev1}.  Such a high
resolution in the measurement of RI using SPR was successfully
utilized for quantification of optical anisotropy in ultrathin
films. Anisotropy in thin film arises due to tilt of shape anisotropic
molecules (e.g. rod shaped calamitic liquid crystal molecules) with
respect to surface normal which may yield in-plane nematic ordering.
In an earlier report by our group, the optical anisotropy in ultrathin
films was estimated experimentally using the SPR phenomenon by
measuring shift in the RA in orthogonal directions of the films
exhibiting different degree of optical anisotropy~\cite{devJMS}.  The
reported anisotropy in the ultrathin films was estimated from SPR
angle measurements in randomly chosen orthogonal directions.  In order
to estimate the in-plane birefringence ($\Delta{n}=n_{e}-n_{o}$), the
SPR measurement has to be performed along the optic axis (OA) of the
thin film and orthogonal to it.  The measured values of RI along OA
and orthogonal to it can be treated as extraordinary ($n_{e}$) and
ordinary ($n_{o}$) components, respectively~\cite{bire}. The shift in
RA along OA and orthogonal to OA of a given anisotropic thin film
exhibiting nematic ordering in two dimensional plane can be defined as
$\Delta\theta$.  In the present work, we have modified
  our experimental setup by integrating a rotating platform (rotation
  axis along X-axis, Fig.~\ref{fig1} ) with a resolution of
  0.1$^{\circ}$ to rotate the film deposited substrate and measure the
  SPR response in-situ as a function of angle of rotation of the
  film. This modification ensures alignment of optics for the
  measurement of $n_e$ and $n_o$ and hence $\Delta{n}$ of the
  ultrathin film.  The RI of ultrathin film will be dependent on
several factors including the surface density, orientation of
molecules, surface morphology and the thickness of the film. Thus the
RA measured using SPR phenomenon will be dependent on such
factors. Therefore, a systematic study is needed for the estimation of
important optical parameter related to thin film viz. in-plane
birefringence ($\Delta{n}$).

The reports in literature in general provide the value of
birefringence of the bulk material however, due to reduction of
dimension of the material, the physical properties deviate largely
from that of bulk. Therefore, measurement of physical properties of a
material at the lower dimension is essential for material engineering
followed by device fabrication. The physical properties of the low
dimensional materials like two dimensional thin film depend on its
thickness.  Hence, a calibration curve is essential for quantifying
the dependencies of a physical property on any such parameters. Since,
the SPR phenomenon can be potentially employed for the measurement of
RI at a very high resolution, a small in-plane birefringence due to
tilt of shape anisotropic organic molecules even in a single layer can
be measured. Such film with tilted molecules may exhibit nematic
ordering on the surface. In this article, we present a calibration
surface showing the dependency of $\Delta\theta$ on $\Delta{n}$ and
thickness of the thin organic film. The calibration surface was
obtained through simulation and it was utilized for the estimation of
$\Delta{n}$ of single layer of Langmuir-Blodgett (LB) films of cadmium
stearate (CdSA) and 4'-octyl-4-biphenylcarbonitrile (8CB)
molecules. The values of thickness and $\Delta\theta$ of the LB films
of CdSA and 8CB was obtained from X-ray reflectivity and a home built
SPR instrument, respectively and these values were used in the
calibration surface for the estimation of the respective $\Delta{n}$.

\section*{Simulation Setup}
\begin{figure}
  \subfigure[]{\includegraphics[width=9cm, clip=true]{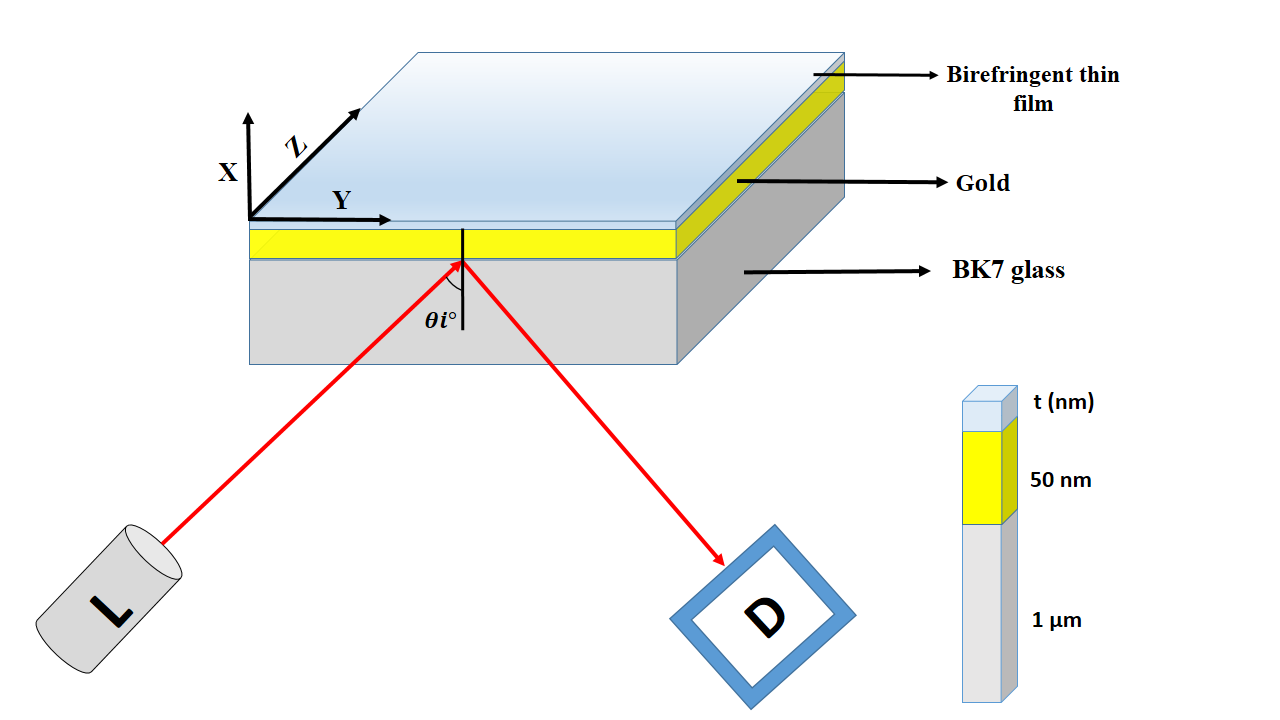}}
  \subfigure[]{\includegraphics[width=4cm, clip=true]{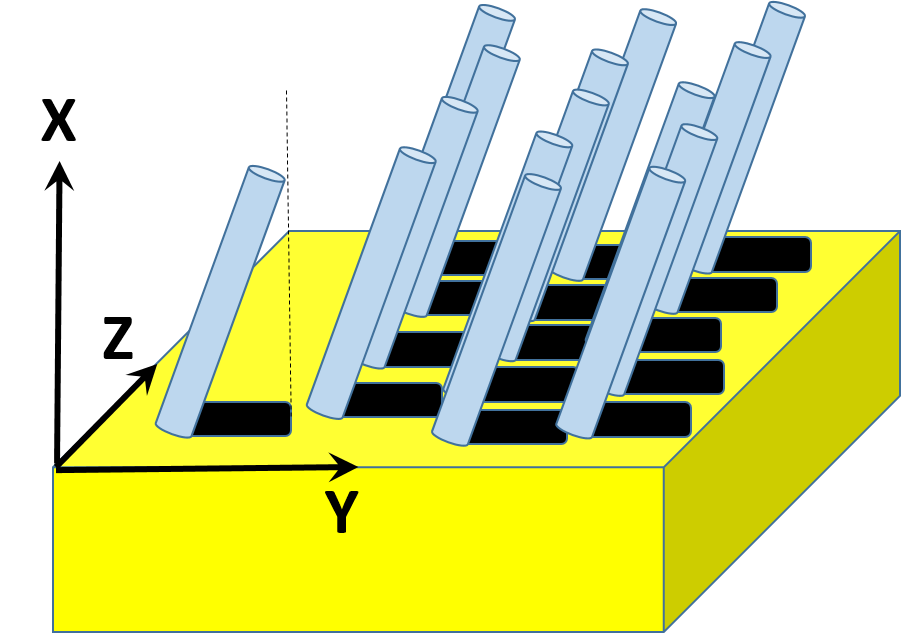}}
  \caption{A schematic of (a) simulation setup showing the major
    components as depicted.  The plane of polarization is XY.  The
    angle of incidence of the monochromatic light (L) is $\theta_{i}$,
    thickness of each material and detector (D) are shown and (b) a
    single layer of shape anisotropic molecules (rod shaped) tilted
    with respect to X-axis along Y-axis on the YZ plane. The
    projection of the molecules is shown in black. Such projection
    resembles nematic ordering on 2D plane with optic axis along
    Y-axis.}
  \label{fig1}
  \end{figure}

A finite difference time domain (FDTD) method was employed for the
simulation of SPR phenomenon in the Kretschmann configuration using a
commercial package of Lumerical~\cite{lum1,lum2}. The FDTD method is
highly reliable and advantageous over other techniques in solving
Maxwell's equations for complex geometries of materials.  The
simulation setup is shown in the Figure~\ref{fig1}(a).
  The simulation was carried out using a monochromatic plane wave
  source (L) having a wavelength of 635 nm. The perfectly matched
  layer (PML) boundary condition with steep angle profile of 12 layers
  was used in order to minimise reflection from the boundary as the
  wave enters into the layer. Linear discrete Fourier transform
  monitors were used to capture reflected and transmitted electric
  field at 350 nm away from the interface. The source was made to
  incident on the gold layer via glass medium at an angle of incidence
  of $\theta_{i}$. In order to obtain the resonance angle, the
  incident angle sweep was generated from 40$^{o}$- 48$^{o}$ with 251
  iterations.  The mesh override was selected in the propagation
  direction of the plane wave to get more precise results. The optical
  anisotropy was seen in case of a single layer of materials
  exhibiting geometrical anisotropy at the molecular level.  A common
  example of such system is shown schematically in
  Figure~\ref{fig1}(b). A single layer of rod shaped molecule
  (calamitic liquid crystal) tilted with respect to X-axis can have a
  projection on the YZ plane. If all the molecules are more or less
  tilted in the same direction (here it is along Y-axis), they exhibit
  a nematic-like ordering with optic axis (OA) parallel to the Y-axis.
  Another set of examples are single layer of self-assembled monolayer
  of rod shaped octadecanethiol or Langmuir-Blodgett film of fatty
  acids~\cite{LBSA}.  To simulate such system of anisotropic material,
  a thin layer of organic material was added onto the gold  layer whose
  in-plane birefringence $(\Delta{n})$ was varied systematically to
  observe the change in the resonance angle  for
  the same system but measured along the OA (i.e. Y-axis) and
  orthogonal to it (i.e. Z-axis) in the SPR simulation model. Since
  the material is organic, only the real part of RI is
  considered in the simulation.

\section*{Experimental}
The Kretschmann configured SPR instrument was developed in the
laboratory~\cite{devSNB}. The equipment utilizes 5 mW laser of
wavelength 635 nm, coupling prism (RI=1.51) and a segmented photodiode
as detector.  The resolution and sensitivity of the equipment are
1.92~$\mu{RIU}$ and 53$^{\circ}/RIU$, respectively. The SPR chip
consists of 0.5 mm glass plate (RI=1.51) deposited with 50 nm thick
gold film through sputtering technique. The chemicals, stearic acid
and 4'-octyl-4-biphenylcarbonitrile (8CB) were procured from
Sigma-Aldrich. Both the molecules yield a very stable Langmuir
monolayer at the air-water interface and are ideal systems for
utilizing them for fundamental studies~\cite{LBSA,8CB}. A single layer
of LB film of CdSA deposited at 30 mN/m can yield an average molecular
tilt of $\sim$10$^{\circ}$ with respect of surface
normal~\cite{cdsatilt} and similarly, that of 8CB deposited at 4 mN/m
yields an average molecular tilt of $\sim$60$^{\circ}$ with respect to
the surface normal~\cite{8cbtilt}.  A single layer of LB films of CdSA
and 8CB were deposited onto SPR chips at target surface pressure of 30
and 4 mN/m, respectively using a LB trough (KSV-NIMA).  The thickness
of the LB films were measured by X-ray reflectivity (XRR) technique
using a X-ray diffractometer equipped with thin film analysis unit
(SmartLab, Rigaku).

\section*{Results and Discussion}

\begin{figure}[htbp]
  \subfigure[]{\includegraphics[width=6.5cm,clip=true]{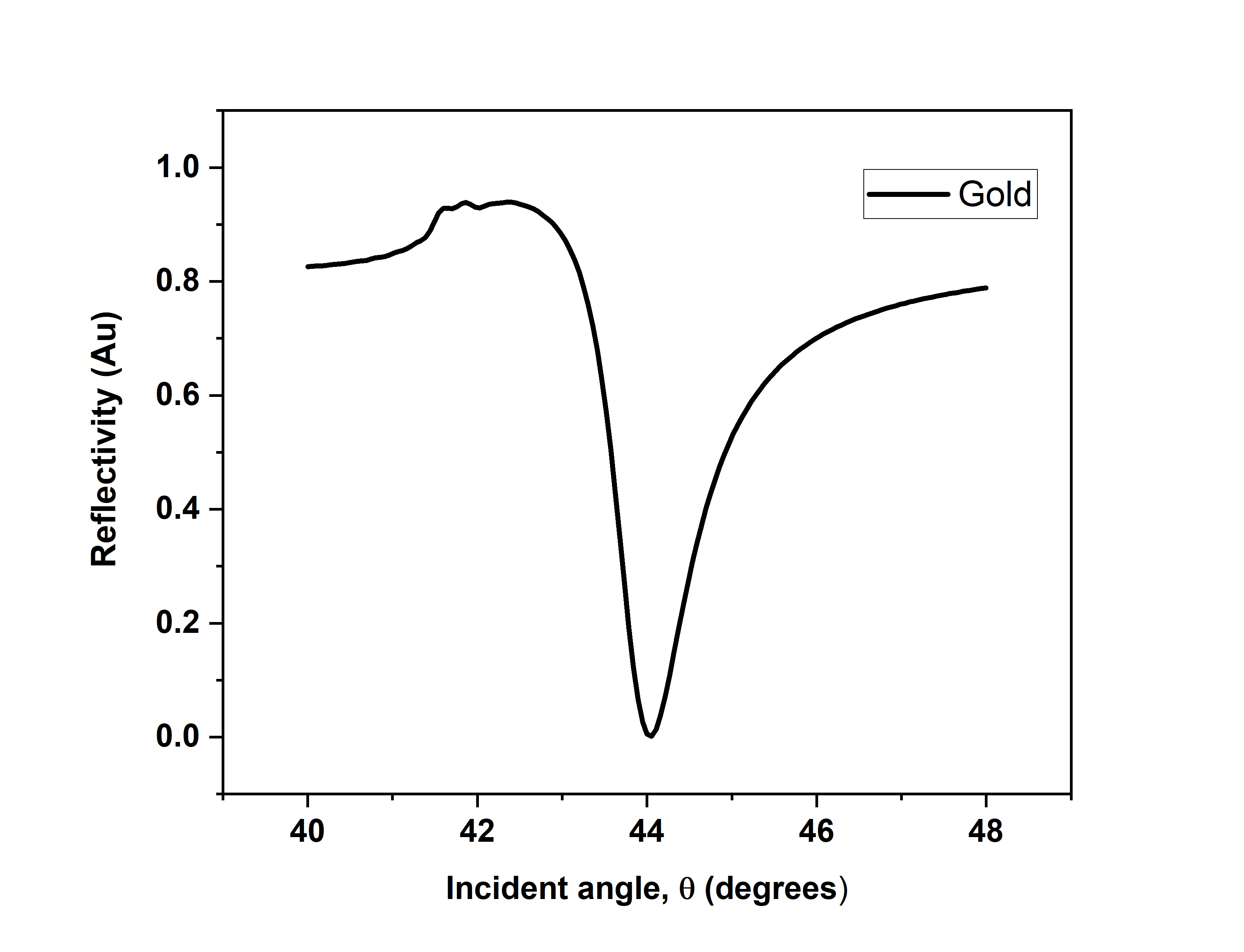}}
  \subfigure[]{\includegraphics[width=6.5cm,clip=true]{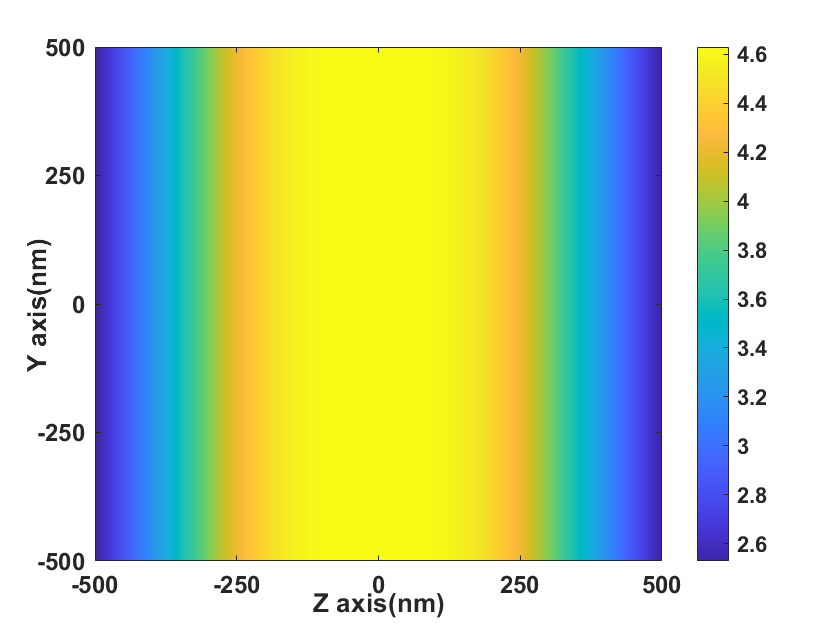}}
	\caption{(a) SPR spectrum of gold-air interface (b) the
          electric field profile on the two dimensional gold surface
          (Y-Z plane) obtained from simulation.}
	\label{fig3}
\end{figure}
A p-polarized electromagnetic wave was allowed to incident at the
glass-gold interface as shown in the Figure~\ref{fig1}. The evanescent
wave generated in the gold film can excite the surface plasmon
polaritons (SPP).  Figure~\ref{fig3}(a) shows the SPR curve for the
gold-air interface.  It exhibits the RA value of 44$^{\circ}$. The SPR
curve and hence the RA value obtained through the FDTD calculation for
the gold-air interface is in agreement with the
literature~\cite{Gupta2017}. The two dimensional (2D) electric field
profile due to the surface plasmon polaritons at the resonance angle
was obtained and is shown in Figure~\ref{fig3}(b). According to the
chosen geometry, the YZ plane corresponds to the gold-air interface
and the plane of polarization is XY.   The SPP are
  excited by the incident p-polarized electromagnetic wave. Therefore, the
  electric field of the incident electromagnetic wave is restricted in
  the XY plane and has zero component along the Z-axis.  This may
lead to surface distribution of the surface plasmon field to be
anisotropic in nature.  For a chosen 1000 nm$\times$1000 nm mesh size,
the anisotropic nature of the plasmonic field can be clearly seen in
the image. This indicates that the excitation of SPP is non-isotropic
and hence there is an immense possibility that coupling of such
anisotropic field with optically anisotropic material will be
direction dependent. Therefore, the SPR measurement of such
anisotropic materials in different direction with reference to the
plane of incidence can yield different resonance angle. The materials
with optical anisotropy can be obtained either in bulk state or as a
single layers of organic molecules exhibiting some shape
anisotropy. The rod shaped calamitic liquid crystal molecules exhibit
a birefringence of $\sim$0.2 in the bulk nematic
phase~\cite{LCbire,5CBbire}. The liquid crystal molecules have great
technological importance where such optical anisotropy play
significant role in display device applications. When such shape
anisotropic molecules are aligned onto solid substrate through
self-assembly or a controlled Langmuir-Blodgett deposition
technique~\cite{roberts}, the deposited single layer can induce a
degree of optical anisotropy due to a collective tilt of the molecules
with respect to the surface normal.  Hence the projections of such
tilted molecules can yield a nematic ordering on the two dimensional
surface.  In our simulation setup, we created an organic layer of a
given thickness whose RI is chosen to be anisotropic by assigning
different values along X, Y and Z axes. The SPR spectra were obtained
through simulation when the plane of incidence is parallel and
perpendicular to the OA of the in-plane nematic ordering in thin film
of organic material. The difference in RA was noted as $\Delta\theta$
from the SPR spectra obtained in these two geometries.

\begin{figure}[htb]
	\includegraphics[width=8cm,clip=true]{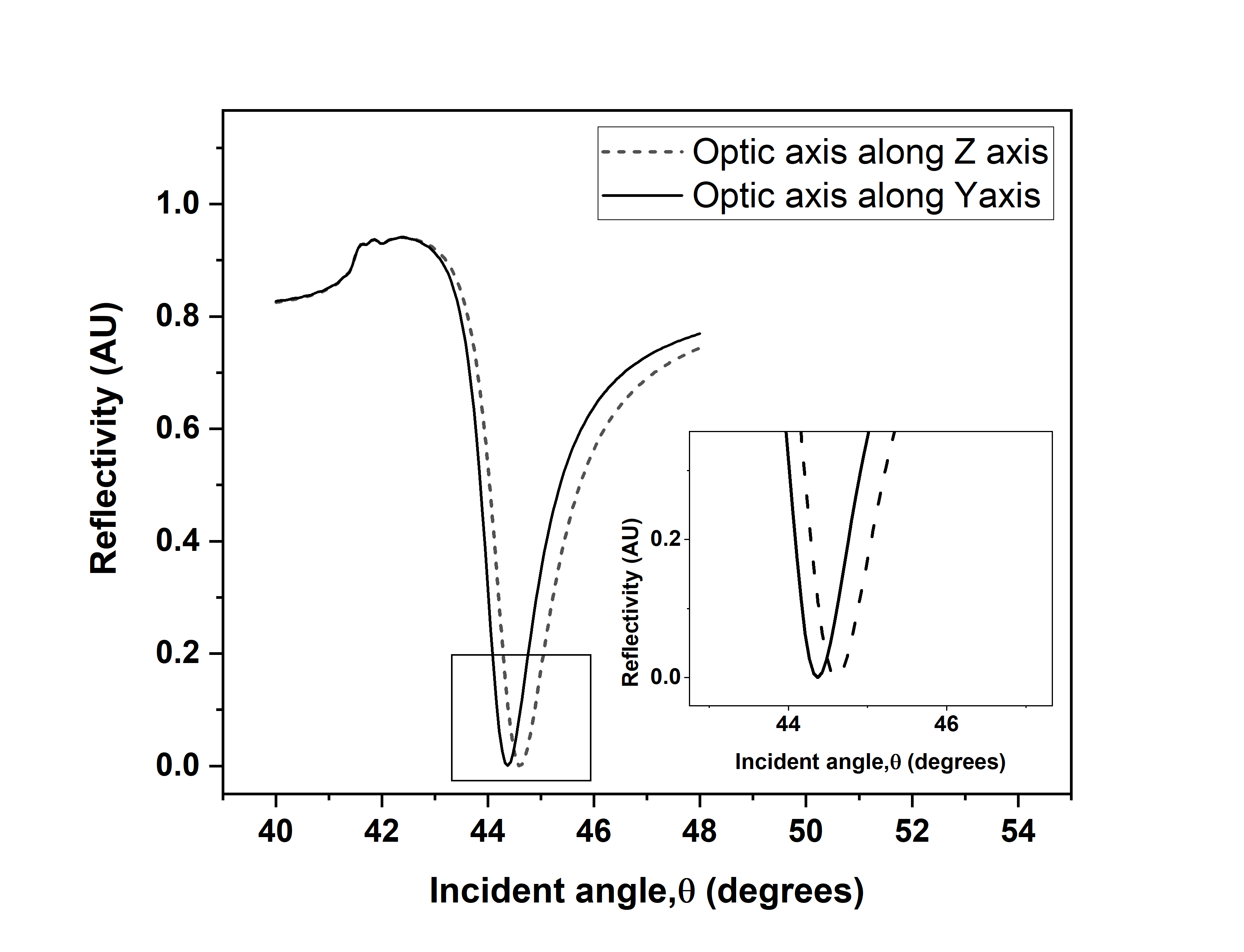}
	\caption{Simulated SPR spectra of a 2 nm thick organic film
          consisting of shape anisotropic organic molecules exhibiting an
          in-plane birefringence of 0.1.}
	\label{fig4}
\end{figure}

\begin{figure}
\includegraphics[width=8cm,clip=true]{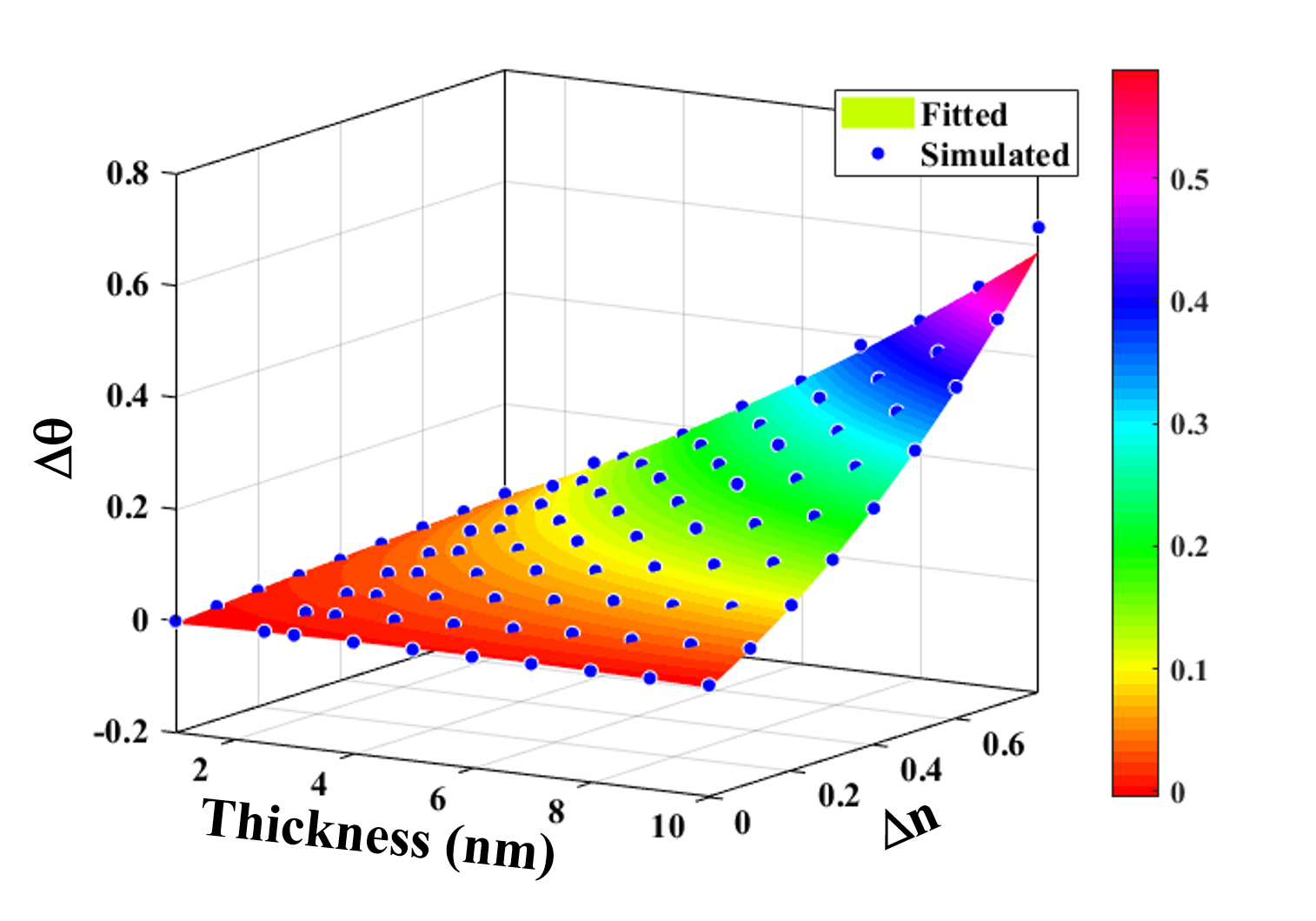}
        \caption{Calibration surface plot showing the variation of
          $\Delta\theta$ as a function of in-plane birefringence
          ($\Delta{n}$) and thickness of organic film. The simulated
          points are shown as filled circle. The surface is
          polynomially fitted. }
		\label{fig5}
              \end{figure}

Figure ~\ref{fig4} shows the SPR curves obtained for an anisotropic
thin film of 2 nm thickness having $\Delta{n}$ as 0.1. The
corresponding RAs were obtained as $44.45^{\circ}$ and
$44.80^{\circ}$ yielding $\Delta\theta$ to be $0.35^{\circ}$.
In the simulation, the SPR curves are obtained for different values of
$\Delta{n}$ and thickness of organic film and the corresponding
$\Delta\theta$ were obtained.  A calibration surface displaying the
variation of $\Delta\theta$ as a function of $\Delta{n}$ and film
thickness ($t$) is plotted in Figure~\ref{fig5}.
              The simulated data are fitted with a surface polynomial curve
              \begin{equation}
                \Delta\theta = P_{1}+P_{2} t+P_{3}\Delta{n}+P_{4} t^{2}+P_{5}\Delta{n}^{2}+P_{6} t \Delta{n}+P_{7} t^{2}\Delta{n}+P_{8}t \Delta{n}^{2}+P_{9}\Delta{n}^{3}
                \label{eqn1}
                 \end{equation}
                 where $P_{i}, i=1,2,3...9$ are the fit
                 parameters. The fit indicator R-square was 0.993
                 which suggests a good fitting. The fitted calibration
                 surface as represented by the Eq.~\ref{eqn1} can be
                 useful for the determination of $\Delta{n}$ of thin
                 films using SPR phenomenon in the very simple
                 prescribed methodology as discussed here.

\begin{figure}[htpb]
      \subfigure{\includegraphics[clip=true,width=5cm]{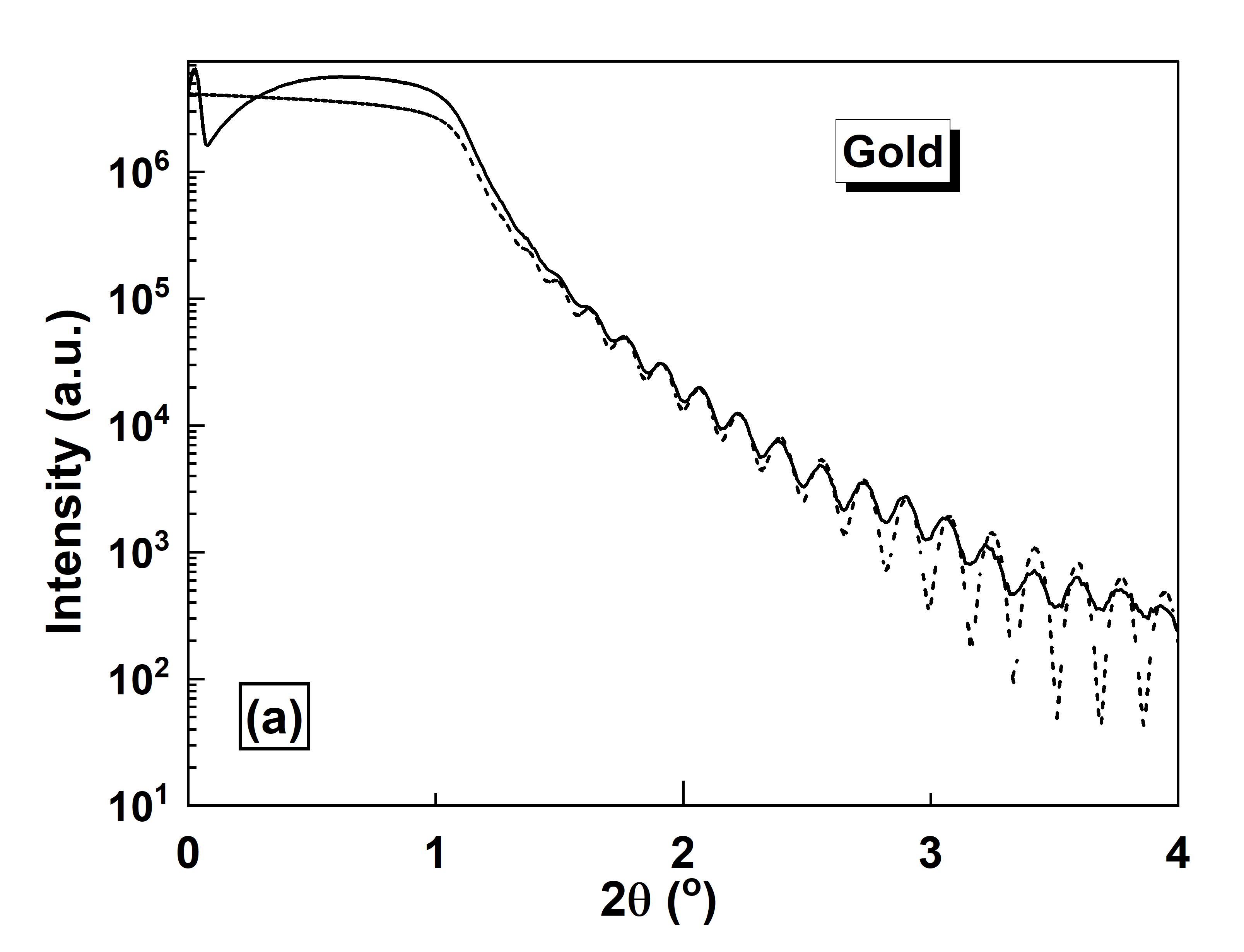}}
      \subfigure{\includegraphics[clip=true,width=5cm]{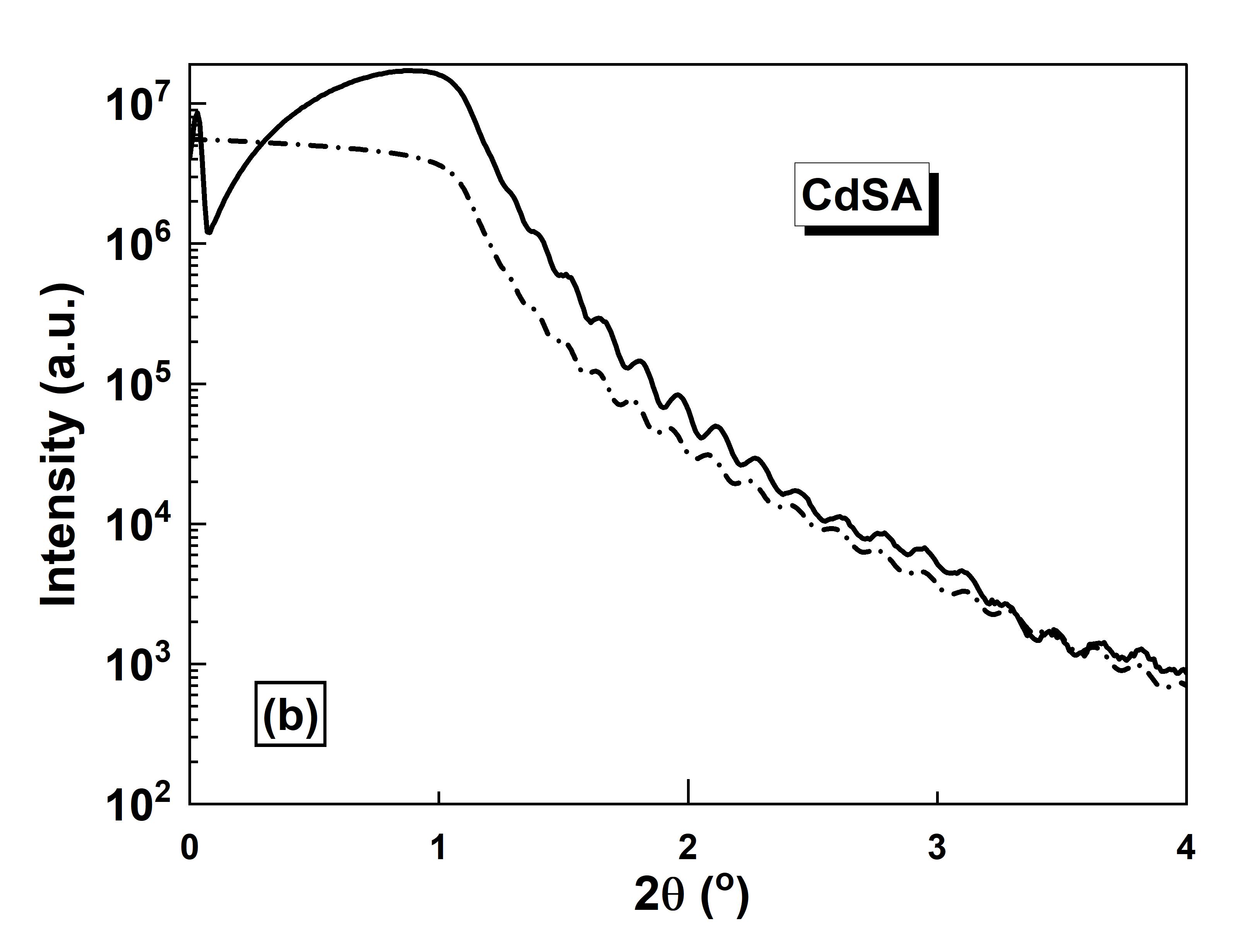}}
      \subfigure{\includegraphics[clip=true,width=5cm]{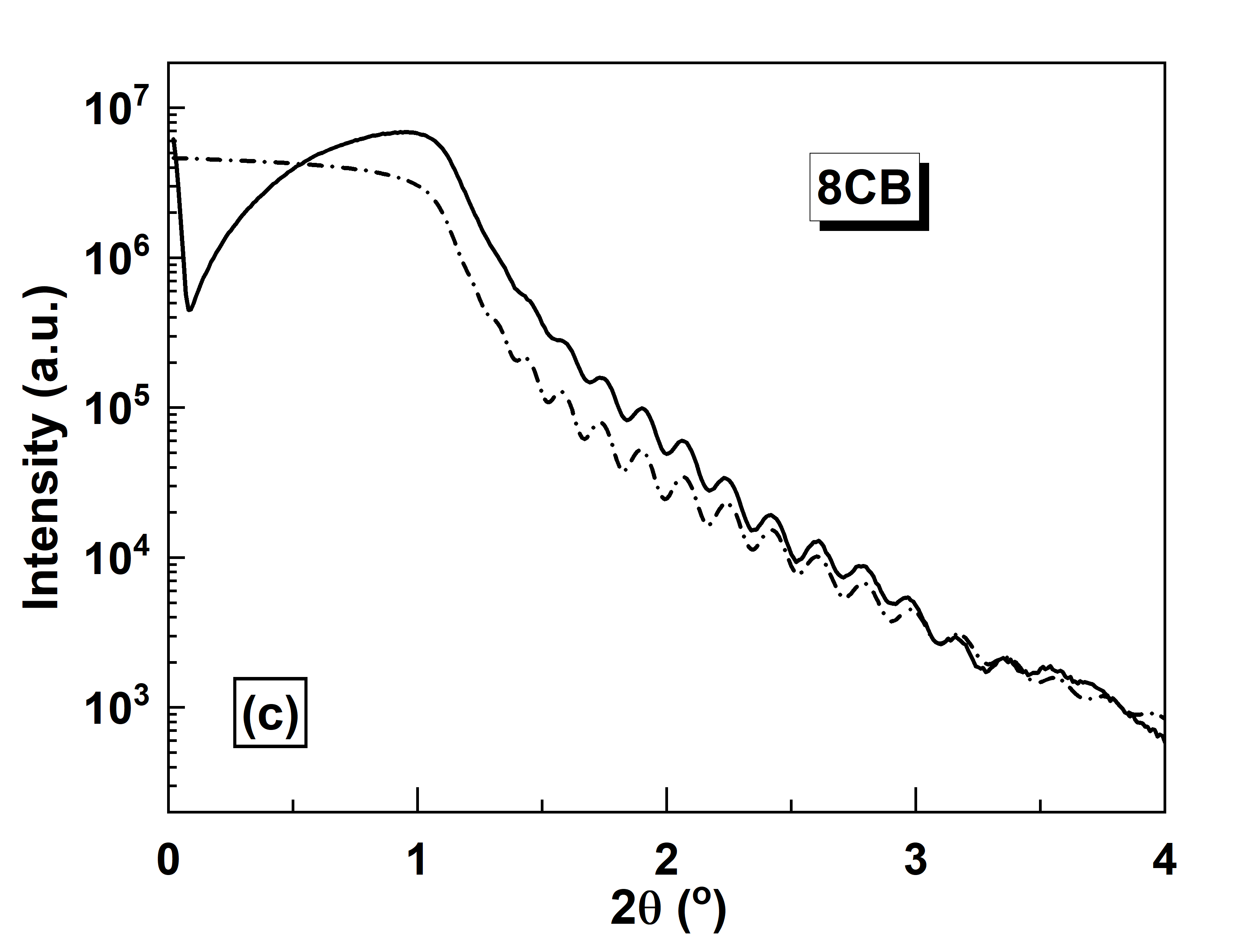}} 
                  \caption{X-ray reflectivity curves obtained from (a)
                     thin films of gold, (b) LB films of cadmium stearate
                     (CdSA) and (c) 4'-octyl-4-biphenylcarbonitrile
                     (8CB). The theoretical fitting yields the
                     thickness of gold, CdSA and 8CB films to be 49,
                     2.4 and 2.0 nm, respectively.}
                   \label{xrr}
                 \end{figure}

                 We have utilized the calibration surface
                 (Eqn.~\ref{eqn1}) for the estimation of in-plane
                 birefringence of ultrathin films fabricated using the
                 standard Langmuir-Blodgett (LB) technique.  We
                 fabricated a single layer of LB films of cadmium
                 stearate (CdSA) and 8CB molecules
                 on the SPR chips at the target surface pressure of 30
                 and 4 mN/m, respectively~\cite{LBSA,8CB}.  The
                 molecules in a single layer of LB films of CdSA and
                 8CB were tilted by $\sim$10 and 60$^{\circ}$ with
                 respect to the substrate
                 normal~\cite{cdsatilt,8cbtilt}. Hence, they can offer
                 anisotropy in the refractive indices and therefore can exhibit
                 non-zero values of $\Delta{n}$. The thickness of the
                 LB films were obtained from X-ray reflectivity
                 measurement (Figure~\ref{xrr}). The experimental
                 curve was fitted using Parrat's
                 formulation~\cite{Parratt} and the thickness of the
                 film was estimated therefrom. The thickness of gold
                 film deposited over the glass plate, LB films of CdSA
                 and 8CB deposited over such gold substrates were
                 estimated as 49, 2.4 and 2.0 nm, respectively.

                 The LB films of CdSA and 8CB were scanned using the
                 SPR instrument. The change in RA
                 along the such orthogonal directions ($\Delta\theta$)
                 were found to be 24 and 71 millidegree, respectively.
                 Such non-zero values suggest the anisotropy in the
                 ultrathin films. The values of thickness and
                 $\Delta\theta$ were substituted in the calibration
                 surface and  $\Delta{n}$ of the
                 ultrathin films of CdSA and 8CB were estimated as
                 0.012 and 0.022, respectively. 

                 Our analysis give a strong foundation for the
                 measurement of in-plane birefringence of ultrathin
                 films of organic molecules. Such information are
                 essential for the development of optical devices.

                 \section*{Conclusion}

                 The measurement of physical properties at a lower
                 dimension is challenging due to large dependencies of
                 the properties on other parameters e.g. thickness of
                 the thin film, aspect ratio of nanomaterials,
                 morphology etc. In this article, we simulated the SPR
                 phenomenon in Kretschmann configuration to measure
                 the in-plane birefringence of thin organic film. The
                 thin film consists of rod shaped organic molecules
                 tilted on the gold surface and thus exhibited
                 in-plane nematic ordering. We performed simulation to
                 obtain a calibration surface showing the variation
                 of $\Delta\theta$ as a function of $\Delta{n}$ and
                 thickness of the film. Such calibration surface was
                 employed for the estimation of $\Delta{n}$ in single layer of LB
                 films of CdSA and 8CB. This study provides a vital
                 methodology for the measurement of very small value
                 of $\Delta{n}$ even in case of a single layer of
                 ultrathin organic film. Further studies involve the role
                 of percolation in quasi-two dimensional film on the optical properties.

\section*{Acknowledgements}
We are thankful to BITS Pilani for providing Lumerical software. We
are thankful to Department of Science and Technology, India for
providing the XRD facility through FIST programme. Thanks are also due
to DST India for supporting SPR instrument from project
(IDP/SEN/06/2015) and LB trough from (CRG/2018/000755).  This is a
post-peer-review, pre-copyedit version of an article published in
Plasmonics. The final authenticated version is available
online at: {\url{https://doi.org/10.1007/s11468-021-01373-1}.

  \section*{Funding}
  Not applicable.

  \section*{Conflicts of interest/Competing interests}
  There are no conflicts of interest/competing interests to declare.

  \section*{Availability of data and material}
  The datasets generated during and/or analysed during the current
  study are available from the corresponding author on reasonable
  request.

  \section*{Code availability}
  Lumerical is GUI based commercial simulation package.
  As such code availability is not applicable. However, some
  scripts can be made available on reasonable request
  to corresponding author.
  
  \section*{Authors' contributions}
  Simulation and part of experiments were done by Amrit
  Kumar. Conceptualization, data analysis, manuscript preparation were
  done by Raj Kumar Gupta. Data analysis and manuscript preparation
  were done by Manjuladevi. SPR measurements and part of experiments
  were done by Ashutosh Joshi.

\section*{References}


\end{document}